\documentclass[aps, reprint, amsmath, amssymb, amsfonts, floatfix, intlimits,superscriptaddress,
showkeys]{revtex4-1}
\pdfoutput=1
\usepackage{graphicx}
\usepackage{dcolumn}
\usepackage{bm}
\usepackage{color} 
\usepackage{float}

\usepackage{url}
\usepackage{color}
\usepackage[colorlinks = true,
	linkcolor = blue,
	urlcolor  = blue,
	citecolor = blue,
	anchorcolor = blue
	]{hyperref}
	
\begin{document}

\title{Impact of geometry and non-idealities on electron `optics' based graphene p-n junction devices}
\author{Mirza M. Elahi}
	\email{me5vp@virginia.edu}
	\affiliation{
		Department of Electrical and Computer Engineering, University of Virginia, Charlottesville, VA 22904, USA\looseness=-1}
\author{K. M. Masum Habib}
	\altaffiliation[Present address: ]{Intel Corp., Santa Clara, CA 95054, USA}
	\affiliation{
		Department of Electrical and Computer Engineering, University of Virginia, Charlottesville, VA 22904, USA\looseness=-1}
\author{Ke Wang}
	\affiliation{
		Department of Physics, Harvard University, Cambridge, MA 02138, USA\looseness=-1}
	\affiliation{
	    School of Physics and Astronomy, University of Minnesota, Minneapolis, MN 55455, USA\looseness=-1}
\author{Gil-Ho Lee}
	\affiliation{
		Department of Physics, Harvard University, Cambridge, MA 02138, USA\looseness=-1}
	\affiliation{
	    Department of Physics, Pohang University of Science and Technology, Pohang 37673, South Korea\looseness=-1}
\author{Philip Kim}
	\affiliation{
	    Department of Physics, Harvard University, Cambridge, MA 02138, USA\looseness=-1}
\author{Avik W. Ghosh}
	\affiliation{
			Department of Electrical and Computer Engineering, University of Virginia, Charlottesville, VA 22904, USA\looseness=-1}
	\affiliation{
	    Department of Physics, University of Virginia, Charlottesville, VA 22904, USA\looseness=-1}
\date{\today}
\begin{abstract}
	We articulate the challenges and opportunities of unconventional devices using the photon like flow of electrons in graphene, such as Graphene Klein Tunnel (GKT) transistors. 
	The underlying physics is the employment of momentum rather than energy filtering to engineer a gate tunable transport gap in a 
	2D Dirac cone bandstructure. In the ballistic limit, we get a 
	clean tunable gap that implies subthermal switching voltages 
	below the Boltzmann limit, while maintaining a high saturating 
	current in the output characteristic. In realistic structures, detailed numerical simulations and experiments show that 
	momentum scattering, especially from the edges, bleeds leakage paths 
	into the transport gap and turns it into a pseudogap. We quantify the 
	importance of reducing edge roughness and overall geometry on the low-bias transfer characteristics of GKT transistors and benchmark 
	against experimental data. We find that geometry plays a 
	critical role in determining the performance of electron optics 
	based devices that utilize angular resolution of electrons.
	\end{abstract}
\maketitle
    In recent years, there has been a number of proposals 
    \cite{williams2011gate, sajjad2011high, jang2013graphene, 
    sajjad2013manipulating, wilmart2014klein, morikawa2017dirac, tan2017graphene} of graphene devices that
    rely on transport gaps \cite{ghosh2015transmission} instead of bandgaps
    exploiting the unique properties of Dirac cone systems
    at p-n junctions. Some of these initial device ideas relied on 
    negative refractive index and Veselago lensing resulting from 
    the conservation of transverse quasi-momentum at the junction
    \cite{cheianov2007focusing, lee2015observation}. 
    However, the switching properties of such waveguide-like devices 
    are likely to be very modest, even for perfect geometries in scaled devices
    \cite{low2009conductance, williams2011gate}, due to the need for sharp injectors and detectors. 
    Angle dependent transmission
    of Dirac fermions \cite{cheianov2006selective} in graphene p-n 
    junction (GPNJ), on the other hand, potentially offers more 
    robust solutions with macroscopic gates and contacts. 
      
\begin{figure}[h!]
        \vspace{-3mm}
    	\includegraphics[width=0.9\linewidth]
    	    {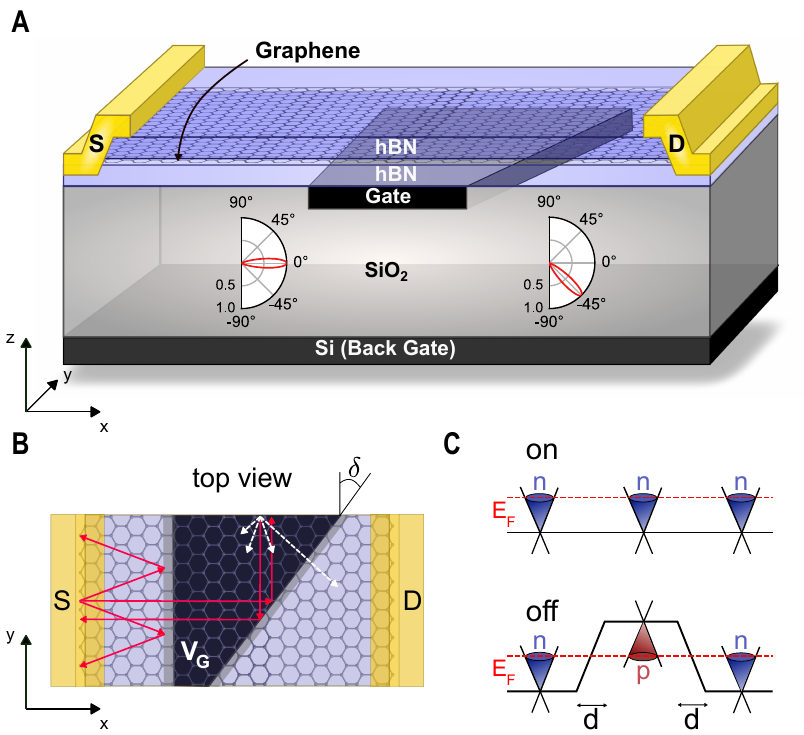}
    	\vspace{-3mm}
    	\caption{
    		\label{fig:device}
    	    \textbf{Graphene Klein Tunnel transistor using
    	   electron optics.} 
    	   (\textbf{A}) 3D schematic. The polar plots in inset show angle dependent transmission
    	    probability of electrons at each junction in the off
    	    state (n-p-n). 
    	    First junction only permits normal incident electrons. 
    	    Second junction, tilted
    	    at $|\delta|=45^\circ$, is allowing only
    	    electrons close to -45$^\circ$, thereby filtering
    	    most of the electrons.
    	    (\textbf{B}) Top view. 
    	    Off state electron paths are shown in red color 
    	    and white color path shows non-specular reflection from 
    	    rough edge resulting in leakage in off state. 
    	    (\textbf{C}) Potential profile in on (n-n-n) 
    	    and off (n-p-n) state. Here, $d$ is the junction width.} 
        \end{figure}

    A perfect match of the pseudospin 
    structure at the interface causes a GPNJ to become completely transparent 
    to normally incident electrons (Klein tunneling 
    \cite{katsnelson2006chiral, young2009quantum}) while it 
    becomes more opaque as the incident angle increases.
    Ramping up the voltage barrier across 
    the junction collimates the electrons by narrowing the distribution 
    of their transmission angles. This collimation can be further enhanced with 
    a smoothly varying barrier of finite width 
    spanning a split gated junction, 
    which imposes an added Gaussian distribution around normal incidence 
    \cite{cheianov2006selective}. Subsequently, putting a second 
    junction at a relative angle ($\delta$) rejects most of the 
    electrons as long as $\delta$ 
    exceeds the maximum critical angle ($\theta_C$) of the filtered and collimated electrons \cite{sajjad2013manipulating}. 
    This two junction device, 
    analogous to a polarizer/analyzer in optics, is broadly 
    referred to as Graphene Klein Tunnel (GKT) transistor (Fig. \ref{fig:device}).
    
    Angle dependent transmission is key to getting a tunable resistance
    in a GKT, achieved by controlling the gate voltage.
    Sajjad \textit{et al.} have shown that such a GKT
    transistor would show a clean transport gap in the off state leading to a nearly ideal 
    transfer 
    characteristic 
    consisting 
    of low off 
    current, 
    high 
    on-off ratio 
    ($I_{on}/I_{off}$=$R_{off}/R_{on}$=10$^4$) 
    and steep subthreshold swing (SS) lower than the 
    Boltzmann limit of 60 mV/decade
    \cite{sajjad2011high, sajjad2013manipulating}. 
    Beyond a desirable gate transfer
    characteristic,
    the GKT transistor was also shown to have 
    an excellent output characteristic
    with a high saturating on current retaining a high mobility in 
    the on state \cite{elahi2016current, tan2017graphene}. 
    In these calculations \cite{sajjad2011high, sajjad2013manipulating, elahi2016current} however, non-idealities
    such as momentum scattering, in particular at the edges were not considered.

    Edge scattering of rejected electrons or holes at the 
    second junction compromises the off state leakage current, 
    as the charge carriers keep bouncing around until some of them 
    find themselves in the narrow transmission lobe of the second junction.
    Indeed, considering edges and
    secondary bounces, 
    a more realistic calculation using both quantum and 
    semi-classical models showed that the on-off ratio
    degrades to $\sim$10$^2$ for perfect edges at widths of
    $\sim$1 $\mu$m \cite{jang2013graphene, wilmart2014klein, elahi2017gate}. 
    Based on the
    initial two junction device idea 
    \cite{jang2013graphene, sajjad2013manipulating}, Morikawa \textit{et al.}
    \cite{morikawa2017dirac} and Wang \textit{et al.} \cite{wangke2018quantum}
    reported experimental on-off ratios 
    of 1.3  and 6-13 respectively, but these on-off ratios are low compared to predictions. 
    Multiple experiments have now confirmed the basic 
    physics of angle dependent transmission at a 
    single tilted junction 
    \cite{sutar2012angle, chen2016electron}, 
    and impact of Klein tunneling in a graphene quantum dot
    \cite{gutierrez2016klein},
    yet no rigorous 
    study has been found 
    explaining the poor on-off ratio in double 
    junction devices in general.
    
\begin{figure}[t]
    	\includegraphics[width=\linewidth]
    	    {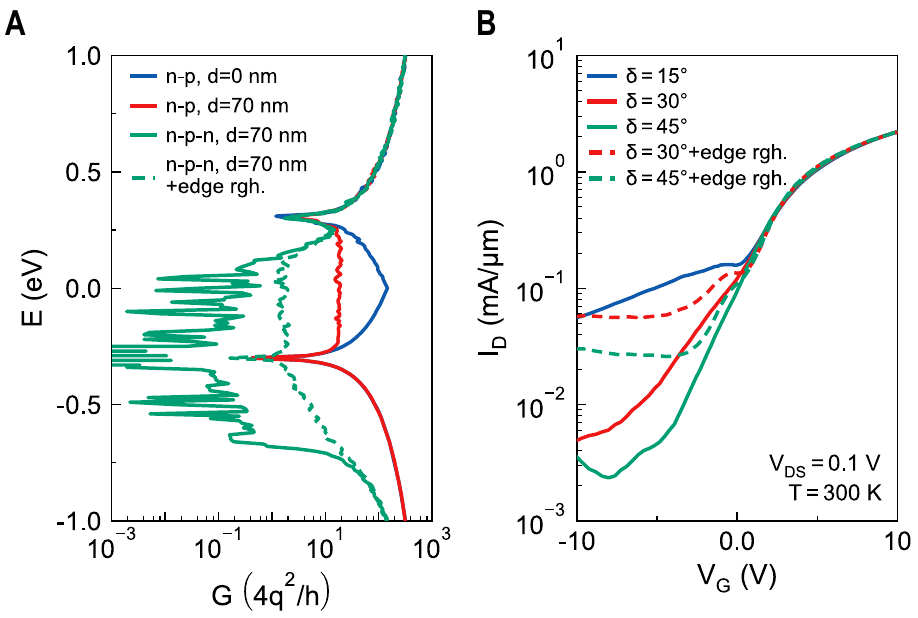}
    	\vspace{-7mm}
    	\caption{
    		\label{fig:Summary} 
    	    \textbf{DJT Device characteristics.} 
    	    (\textbf{A}) Conductance of p-n junction devices 
    	    in off state (n-p or n-p-n). 
    	    Transport gap between -0.67 eV and +0.3 eV arises 
    	    due to electron filtering in ideal n-p-n device 
    	    with $|\delta|$=45$^\circ$. Adding edge roughness increases 
    	    the floor value of the gap shown by dashed line. For comparison 
    	    we also show single n-p junction conductance 
    	    (abrupt and smooth with junction width $d$ = 70 nm). 
    	    (\textbf{B}) Transfer characteristics from 
    	    semiclassical ray tracing simulation with source-drain voltage $V_{DS}$=0.1 V.}
    	    \vspace{-7mm}
        \end{figure}

\begin{figure}[t]
    \centering
	\includegraphics[width=\linewidth]
	    {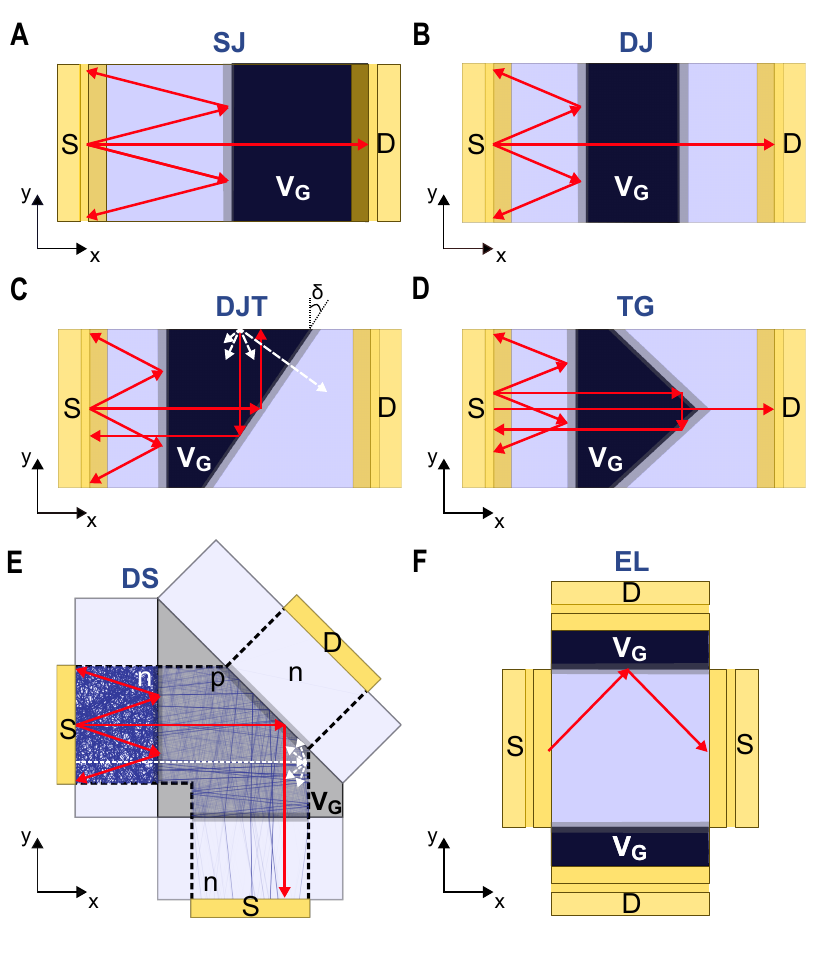}
	\vspace{-6mm}
	\caption{\label{fig:deviceFamily} 
	    \textbf{Device family schematic.} 
	    (\textbf{A}) SJ device. 
	    (\textbf{B}) DJ device. 
	    Both the junctions are parallel to each other. 
	    (\textbf{C}) DJT device with $\delta$=45$^\circ$. 
	    White arrows show spreading of 
	    electrons when they hit the edge in case of non-specular 
	    reflections and leakage path through the second junction 
	    (small incidence angle). 
	    (\textbf{D}) TG device. 
	    Only gate is used to reflect back the electrons 
	    making the device less susceptible to edge roughness, 
	    although tip of the triangle suffers from leakage path. 
	    (\textbf{E}) DS device
	    \cite{wangke2018quantum} with ray tracing simulation paths. 
	    Electrons are reflected back to the other source, 
	    thus it is free from multiple bounce issue of electrons. 
	    (\textbf{F}) EL device. 
	    The junction is rotated 90$^\circ$ with respect to source, 
	    thus most of the electrons are reflected back in off state. 
	    Moreover, this device does not have an edge so edge 
	    roughness does not play any role in this device's performance. 
	    In all the cases, red arrows show electrons path in off state.}
	    \vspace{-5mm}
    \end{figure}

    In this paper, we explain the existing 
    discrepancy between simulations (on-off ratio $\sim$10$^2$)
    \cite{jang2013graphene, wilmart2014klein, elahi2017gate} and 
    experiments (on-off ratio $\sim$10) 
    \cite{wangke2018quantum} of GKT
    devices.
    We find that in addition to the electrons suffering multiple bounces around the wedge shaped region between junctions, 
    non-specular (diffusive) scattering by rough edges, shown by white arrow in Fig. \ref{fig:device}B, 
    plays an important role in degrading the on-off ratio 
    by transforming the transport gap to a pseudogap 
    with a non-zero floor (Fig. \ref{fig:Summary}A).
    We study several variations of graphene p-n 
    junction based devices. 
    Starting from a basic building block, a single p-n junction, 
    we extend to complex structures consisting 
    of multiple junctions. 
   Specifically, we consider a split-gated single junction (SJ), 
    parallel split-gated dual junctions (DJ), 
    two split-gated dual junctions in an angled 
    trapezoidal geometry (DJT) \cite{ sajjad2013manipulating}, 
    a triangular gated (TG) prism geometry 
    (basic building block of saw-tooth geometry 
    \cite{ jang2013graphene, wilmart2014klein}), 
    a dual-source (DS) device \cite{ wangke2018quantum}, 
    and finally a structure with two drains rotated at 
    90$^\circ$ relative to two sources, generating 
    an effective EdgeLess device (EL). 
    In Fig. \ref{fig:deviceFamily}, all the structures 
    are shown with their off state electron paths marked in red.
    Our comparative study shows advantages and 
    disadvantages of one structure over another, 
    providing a guideline for designing 
    electron optics inspired devices in future.
    
    A finite transport gap generated by the angular filtering of 
    electrons differentiates GKT devices from 
   conventional graphene Field Effect Transistors (gFET). 
    In Fig. \ref{fig:Summary}A, we see a 
    transport gap 
    arising from the double junction structure (DJT). 
    We also show the case for 
    abrupt p-n junction ($d$ = 0 nm) where 
    filtering is not that robust. 
    A smooth p-n junction ($d$ = 70 nm) 
    performs better than an abrupt one due to added 
    Gaussian filtering due to angle-dependent tunneling. 
    In presence of edge roughness, 
    the transport gap turns into a gap with 
    a nonzero floor and increases the overall off state conductance
    (Fig \ref{fig:Summary}A). We see dips at $E=\pm0.3$ eV due to Dirac points.
    As shown in Fig \ref{fig:Summary}B, edge roughness degrades the off state performance ($V_G \sim-$10 V) for any given $\delta$. We also show that 
    $\delta$ = 45$^\circ$ gives the lowest off current
    even in the presence of edge roughness, as suggested earlier \cite{sajjad2012manifestation}. Here, local gate dielectric 
    (hexagonal boron nitride, hBN) thickness is 32 nm. To discuss the effect of edge roughness 
    in detail as well as the dependence on device geometry, 
    we analyze a variety of structures in this paper 
    (Fig. \ref{fig:deviceFamily}). 
    
    In this study, we adopt semiclassical ray tracing approach 
    \cite{chen2016electron, elahi2017gate} based on a billiard model 
    \cite{beenakker1989billiard, milovanovic2013spectroscopy, milovanovic2014magnetic} 
    that has been benchmarked 
    against experiments \cite{chen2016electron}. 
    A charge carrier hitting a perfect edge reflects 
    back with an angle equal to the incident angle (specular reflection). 
    In presence of edge roughness, a Gaussian distributed random angle of reflection with standard deviation $\sigma_e$ (higher $\sigma_e$ denoting rougher edges) 
    is added.
    The transmission probability ($T$) for each electron across 
    a junction is calculated analytically, using a generalized version \cite{sajjad2013manipulating} (eqn.~S1 \cite{SI2018}) of the well-known  equation \cite{cheianov2006selective} for symmetric junction, $T \sim e^{ -\pi k_F \frac{d}{2} \sin^2 \theta}$. Here, $k_F$ is the magnitude of the Fermi wave vector on each 
    side for a symmetric p-n junction, $d$ is the junction width, and $\theta$ is the incident angle at the junction. 
    We calculate channel resistance $R_{Ch}$
    for low-bias and total resistance using 
    $R_T = R_{Ch} + 2R_C$, where $R_C$ is the contact resistance between graphene and source/drain electrodes \cite{SI2018}. To explain experimental data
    \cite{wangke2018quantum}, contact resistance $R_C\sim$ 100 $\Omega$-$\mu$m 
    and non-specular edge scattering are included 
    in our semiclassical simulation model. 
    The local gate dielectric (hBN) thickness is 32 nm, junction width $d$ is 70 nm, temperature is 50 K, and 
    device width is 1 $\mu$m unless otherwise mentioned. The main advantage of 
    ray tracing over the Non-equilibrium Green’s function (NEGF) formalism
    is its computational practicality.
        
\begin{figure}[t]
    	\includegraphics[width=\linewidth]
    	    {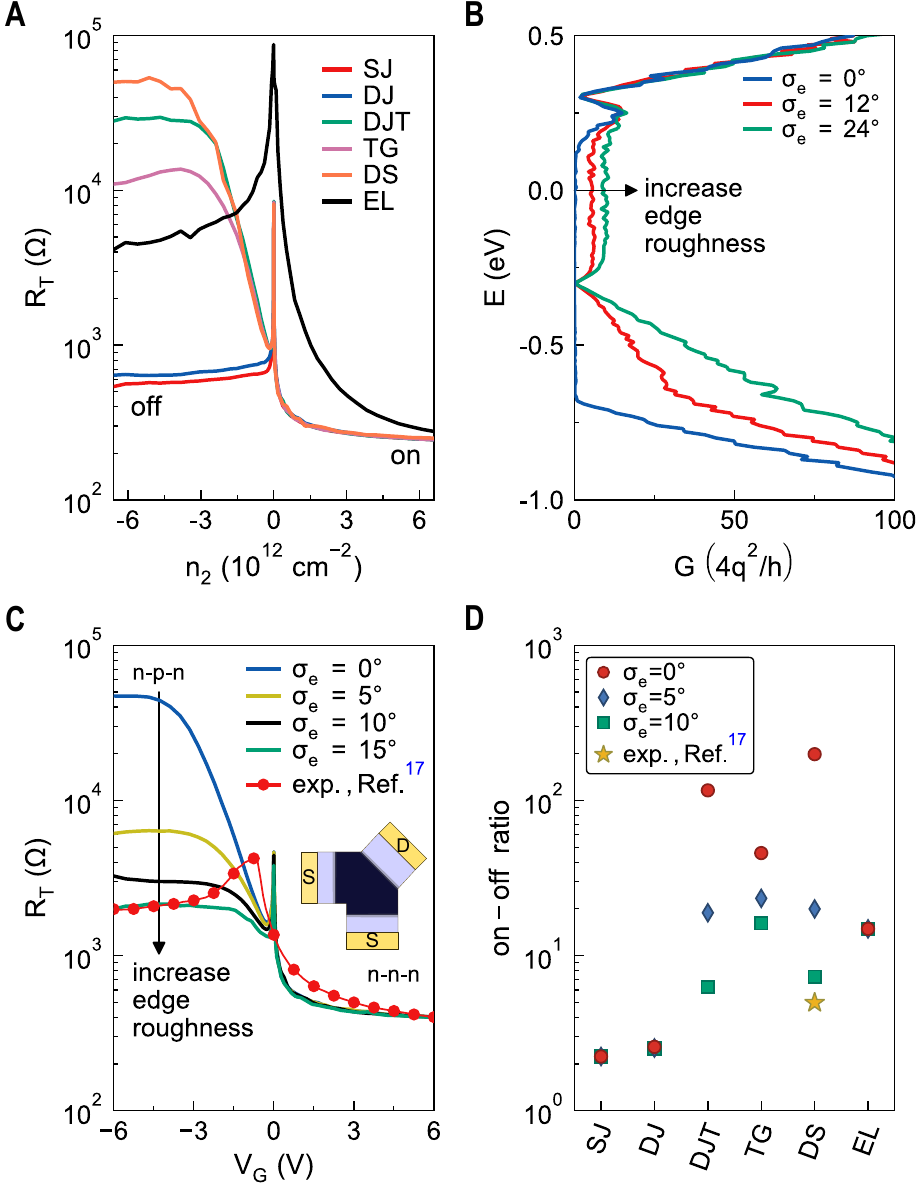}
    	\vspace{-6mm}
    	\caption{\label{fig:EdgeRgh_Panel}
    	    \textbf{Low-bias resistance characteristics and effect of edge
    	    roughness}. 
    	    (\textbf{A}) Calculated total resistance for different geometries.
    	    (\textbf{B}) With increasing edge roughness in a DS device, transport gap turns in to a pseudo gap
    	    having higher nonzero floor value due to additional leakage path.
    	    (\textbf{C}) Low-bias resistance characteristics of 
    	    DS device (experiment \cite{wangke2018quantum} vs. simulation). 
    	    Off state (n-p-n) degrades significantly with 
    	    increasing edge roughness. Contact resistance 
    	    (117 $\Omega$-$\mu$m) and edge roughness parameter
    	    ($\sigma_e$=15$^\circ$) are adjusted to match experimental values of on and off state resistance respectively. 
    	    (\textbf{D}) On-off ratio for different gate geometries with various edge roughness. For reference, experimental on-off ratio 
    	    \cite{wangke2018quantum} from (C) is also shown, 
    	    although device width and doping conditions are not same as
    	    simulation.}
    	    \vspace{-5mm}
        \end{figure}

    We now discuss the impact of gate geometry on 
    various flavors of ballistic, perfect edge GKT transistors, 
    as quantified by their low-bias resistances 
    and on-off ratios.
    Figure \ref{fig:EdgeRgh_Panel}A shows the results of the 
    low-bias on and off state resistances for each geometry. 
    The back gate voltage is kept fixed to $V_{BG}$=100~V 
    (corresponding to charge density $n_1$=6.63$\times$10$^{12}$ cm$^{-2}$ for SiO$_2$ thickness of 300 nm in addition to 32 nm hBN)
    for all these devices while we sweep the local gate $V_G$ 
    to vary the corresponding charge density of middle gate region ($n_2$) 
    from negative (p-type) to positive (n-type), 
    giving us the off and on states respectively. 
    Our first structure, an SJ device 
    (Fig. \ref{fig:deviceFamily}A) filters out 
    carriers at angles other than normal incidence, 
    exhibiting Klein tunneling. 
    Adding another junction aligned to the first one 
    (Fig. \ref{fig:deviceFamily}B, DJ) 
    does not help in increasing on-off ratio significantly, but instead adds another comparable 
    resistance along the path. 
    With a tilted second junction (Fig. \ref{fig:deviceFamily}C, DJT 
    with $\delta=45^\circ$), 
    we can achieve orders 
    of magnitude larger off state resistance for ballistic flow. 
    Next the TG device 
    (Fig. \ref{fig:deviceFamily}D) 
    uses the second junction
    to reflect back strongly collimated carriers
    towards the source away from the edges. 
    However, it has a poorer performance in the off 
    state than DJT structure
    because it allows electrons to Klein tunnel 
    through its vertex on the first try. 
    The DS device \cite{wangke2018quantum} 
    (Fig. \ref{fig:deviceFamily}E) has an overall L-shape, 
    so that each segment of the split source recaptures 
    carriers injected from the other segment and rejected 
    by the tilted junction, without letting them bounce again 
    at the edges. 
    As a result, its off state performance is superior to the
    DJT device. 
    Finally, the EL 
    (Fig. \ref{fig:deviceFamily}F) device capitalizes 
    on a structure that is free from edge effects. 
    In the EL structure shown, electrons enter along 
    one axis from both sources, while the drains are along 
    a perpendicular axis with the gate induced p-n 
    junctions sitting in between. 
    Such EL structures reduce the off current because 
    most electrons incident at the junction are at large angles. Compared to an ideal DJT device, 
    the off state resistance is still low as it uses 
    only one junction. 
    Moreover, the on state current of the EL, 
    determining its device speed, 
    is compromised by the right angle separating 
    source and drain - moving the drain away from the natural `line of sight' of the injected source electrons. The low on current degrades the
    overall on-off ratio of the EL device. 
    
    Edge roughness tends to decrease the 
    on-off ratio of these devices by diffusive 
    scattering of the reflected electrons providing a leakage path to the drain as shown in Fig. \ref{fig:deviceFamily}(C, E) by the white dashed lines. 
    Thus the transport gap turns into a pseudo gap 
    with a finite floor (Fig. \ref{fig:EdgeRgh_Panel}B) 
    with 
    increasing edge roughness. 
    For ideal edges we see a transport gap spanning 
    -0.67 eV to +0.3 eV. 
    With increasing edge roughness, the floor value 
    of the gap also increases (other than at $E$ = -0.3 eV 
    due to a clear Dirac point, which in turn could be washed 
    out by impurity scattering and puddles \cite{sajjad_puddle}), 
    thus increasing off state conductance 
    and decreasing resistance. 
    In Fig. \ref{fig:EdgeRgh_Panel}C, 
    we show the evolution of resistance 
    characteristics of the DS device 
    with increasing edge roughness. 
    Here we use device parameters $d$ = 60 nm, width=800 nm, and voltages
    $V_G = -$6 V to 6 V, $V_{BG}$ = 60 V, 
    emulating a local gate voltage of 6 V, 
    as in the experiment \cite{wangke2018quantum}. 
    In Ref. \cite{wangke2018quantum}, all the regions 
    (n-n-n/n-p-n) are controlled by local gates 
    whereas in our simulation only the middle region 
    is controlled by a local gate ($V_G$) while other regions 
    are controlled by back gate ($V_{BG}$). 
    We match the on state ($V_G$ = 6 V) result by fitting a
    contact resistance ($R_C$ = 117 $\Omega$-$\mu$m) and 
    off state ($V_G = -$6 V) resistance by fitting edge roughness 
    parameter $\sigma_e$=15$^\circ$. We see a mismatch 
    between our simulation and experiment at $V_G\sim0$ V 
    due to charge puddles that average out the 
    Dirac points \cite{sajjad_puddle}.
    
    Figure \ref{fig:EdgeRgh_Panel}D shows the evolution of the
    on-off ratio for all the device geometries 
    in presence of edge roughness. 
    We clearly see orders of magnitude enhancement of on-off ratio in DJT compared 
    to SJ and DJ, but in the presence of edge roughness the
    on-off ratio degrades significantly.
    In contrast, although the TG device starts with a lower 
    on-off ratio due to Klein tunneling at the vertex, 
    it shows robustness against edge roughness as it 
    directs the collimated electrons away from the edges 
    (Fig. \ref{fig:deviceFamily}D). The DS device is supposed to
    perform better even in presence of edge roughness. 
    However, as the device geometry \cite{wangke2018quantum}
    is not optimized considering edge roughness
    (Fig. \ref{fig:deviceFamily}E, second junction position, 
    electrons shown in white arrow hits edge), the on-off ratio 
    degrades significantly. 
    Improved device geometry 
    (DS$_{\textrm{imp}}$) shows at least 2 times better 
    on-off ratio in presence of edge roughness 
    (Fig. S1 \cite{SI2018}). 
    In Fig. \ref{fig:EdgeRgh_Panel}D, we also show the experimental on-off ratio \cite{wangke2018quantum} 
    from Fig. \ref{fig:EdgeRgh_Panel}C, 
    notably with different device width and doping than for simulation.
    Finally, the EL device which starts 
    with an even lower on-off ratio 
    than the TG device, but the on-off 
    ratio remains constant with increasing 
    edge roughness as the electrons do not hit any 
    edge before getting filtered out. 
    
    As a transistor, a ballistic GKT greatly outperforms 
    wave-guided structures based on the Veselago effect. 
    However, even a GKT faces challenges arising from 
    the presence of edges - in particular rough ones, 
    together with contact resistance and finite doping of graphene by metal contacts. 
    A 1 $\mu$m wide structure with perfect edges is 
    predicted to have a gate transfer characteristic 
    with an on-off ratio $\sim$10$^2$, but current technology 
    limits the edge smoothness and degrades the on-off ratio
    to $\sim$10,  demonstrated experimentally \cite{wangke2018quantum}. 
    Such a low on-off ratio is not yet suitable for digital logic. 
    
    The output characteristic, however, bears more promise.
    At high drain bias ($V_{DS}$), a small transport gap 
    (n$^+$-n-n$^+$, on state) at energies far from the equilibrium Fermi level is predicted to produce a strongly
    saturating $I_D$-$V_D$ that is robust  against 
    edge roughness \cite{tan2017graphene}. Even cases with an on-off ratio $\sim$10 can result in 
    an order of magnitude increment in $r_{out}$ (output resistance) without hurting the mobility. 
    Devices like this with high mobility and output resistance can be quite useful 
    for analog RF applications, delivering a high $f_T$ (unity current gain cutoff frequency) and $f_{max}$ (unity power gain cutoff frequency)
    \cite{tan2017graphene}.
    To improve device performance, a superlattice 
    potential may be incorporated into the device to create an anisotropic 
    band structure and create a much more aggressive collimation of electrons 
    \cite{park2008anisotropic, forsythe2018band}. Further improvements may be possible with abrupt junctions if doping can be improved in the first region, so that the reflected electrons at the second junction are no longer stopped by an abrupt first junction on their way to the source (recall that total internal reflection only works one way like a diode, from a denser to a rarer medium). A major factor in determining the overall performance of all these structures is edge roughness, included here as a phenomenological parameter, the standard deviation $\sigma_e$ of a Gaussian angular smear. The relationship between
    $\sigma_e$ and physical roughness parameters, as 
    well as decay rates extracted from
    magnetoconductance measurements is out of scope of this study and will be 
    reported elsewhere.
    
    
    In summary, the performance challenges of a GKT 
    transistor are outlined in this paper. 
    Although junction line-edge roughness \cite{zhou2018atomic} and other scattering events are expected to play a role, we focused here on
    edge roughness that is expected to be the most deleterious to the on-off ratio through momentum redirection. We quantified the role of graphene-edge roughness and attempted to design around it, such as the
    EL device. We analyzed a family of devices and find that TG, DS$_{\textrm{imp}}$ and EL device are 
    less susceptible to edge roughness. We further showed that an angle of 45$^\circ$ between collimator and reflector gives the best performance even in the presence of edge roughness. 
    Our analysis shows that even with geometry optimization the
    on-off ratio may not be enough for scaled digital switching,
    but may still offer advantages for high frequency RF analog applications
    \cite{tan2017graphene}.
    
    This work was supported by Semiconductor Research
    Corporation's (SRC) NRI-INDEX center. The authors want to thank Cory R. Dean for important discussions.
%


\widetext
\clearpage
\begin{center}
	{\large Supplementary Material for}\vspace{5mm}\\
	\textbf{\large Impact of geometry and non-idealities on electron `optics' based graphene p-n junction devices}
\end{center}
\begin{center}
	\begin{small}
		{\textbf{ Mirza M. Elahi,$^{1,a)}$ K. M. Masum Habib,$^{1,b)}$ Ke Wang,$^{2,3}$ Gil-Ho Lee,$^{2,4}$ Philip Kim,$^2$ Avik W. Ghosh$^{1,5}$}}\\ \vspace{2mm}
		$^1$\textit{Department of Electrical and Computer Engineering, University of Virginia, Charlottesville, Virginia 22904, USA.}\\ \vspace{0.2mm}
		$^2$\textit{Department of Physics, Harvard University, Cambridge, MA 02138, USA.}\\ \vspace{0.2mm}
		$^3$\textit{School of Physics and Astronomy, University of Minnesota, Minneapolis, MN 55455, USA.}\\ \vspace{0.2mm} 
		$^4$\textit{Department of Physics, Pohang University of Science and Technology, Pohang 37673, South Korea.}\\ \vspace{0.2mm}
		$^5$\textit{Department of Physics, University of Virginia, Charlottesville, VA 22904, USA.}\\ \vspace{0.2mm} 
		$^{a)}$\textit{Electronic mail: \href{mailto:me5vp@virginia.edu}{me5vp@virginia.edu}}\\ \vspace{0.2mm} 
		$^{b)}$\textit{Present address: Present address:  Intel Corp., Santa Clara, CA 95054, USA.}\\ \vspace{0.2mm} 
	\end{small}
\end{center}
\setcounter{equation}{0}
\setcounter{figure}{0}
\setcounter{table}{0}
\setcounter{page}{1}
\makeatletter
\renewcommand{\theequation}{S\arabic{equation}}
\renewcommand{\thefigure}{S\arabic{figure}}
\renewcommand{\bibnumfmt}[1]{$^{\textrm{S}#1}$}
\renewcommand{\citenumfont}[1]{S#1}
\renewcommand{\thesection}{S\arabic{section}}
\section{Simulation method}
Semiclassical ray tracing model relies on the assumption that the most relevant quantum 
effects in GKT devices manifest during tunneling at the junctions, 
while for large-scale devices with rough edges, interference effects are 
expected to be washed out under the gated regions.  
Accordingly, we throw electrons from a source with random 
injection angles following a cosine distribution function
\cite{molenkamp1990electron}, and evolve each electronic 
trajectory with constant speed $v_F$ (Fermi velocity) and band effective mass, 
$m=(E_F - qV)/v_F^2$ 
following classical trajectories. 
The transmission probability ($T$) of electrons at the 
junction is calculated [\textcolor{blue}{4, 21}] using
\begin{equation}
T(E_F, \theta_1)= 
\begin{cases}
\Theta(\theta_C - \theta_1)\frac{\cos (\theta_1)\cos (\theta_2)}{\cos^2 \left( \frac{\theta_1 + \theta_2}{2}\right)},& \text{for p-p' or n-n'}\\
\Theta(\theta_C - \theta_1)\frac{\cos (\theta_1)\cos (\theta_2)}{\cos^2 \left( \frac{\theta_1 + \theta_2}{2}\right)} \text{exp} \left[ - \pi d \frac{k_{F1}k_{F2}}{k_{F1}+k_{F2}} \sin (\theta_1) \sin (\theta_2)\right],& \text{for p-n' or n-p'}
\end{cases}
\label{eqs1}
\end{equation}
where, $\theta_1$ and $\theta_2$ are the incident and refraction angle, 
$\theta_C$ is the critical angle from Snell's Law, $d$ is the junction width, and $k_{F1}$ and $k_{F2}$ are the Fermi wave vectors 
on the incident and transmitted side respectively.

Assuming non-interacting charge carriers, 
we consider a fraction $T$ of each electron at the junction 
that passes through, while a fraction $1-T$ is reflected back 
to the incident region. The trajectories of the reflected and 
transmitted fractional electrons are allowed to evolve once again 
through multiple such transmission-reflection events until they 
end up either at the source or the drain. 
The average transmission probability
($T_{ij} = N_j/N_{Total}$) from contact $i$ to contact $j$ is 
calculated by counting electrons ($N_j$) that eventually make 
it to the contact $j$ for a given total number
$N_{Total}$ of carriers injected from contact $i$. 
Thereafter the Landauer-B\"{u}ttiker formalism at low-bias 
is used to calculate channel conductance ($G_{Ch}$) 
by summing up the terminal transmissions. 
\begin{equation}
G_{Ch}(E_F) = \frac{4q^2}{h}\int M(E) \overline{T}(E) 
\left( -\frac{\partial f_0}{\partial E} \right) dE
\end{equation}
where $E_F$ is the Fermi energy, 
$M$ is the number of modes, $\overline{T}$ is the sum over 
all transmissions, and $f_0=f(E-E_F)$ 
is the equilibrium Fermi function.

\section{Geometry improvement of Double Source (DS) device}
\begin{figure*}[h!]
	\includegraphics[width=0.8\linewidth]
	{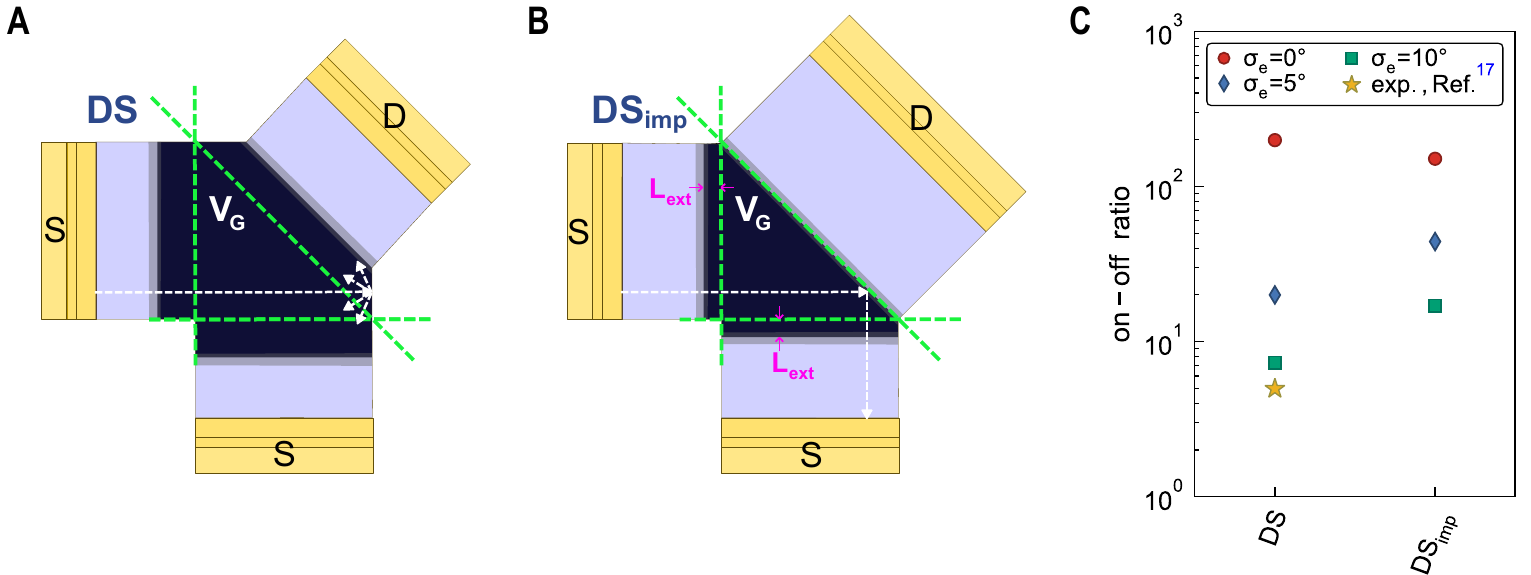}
	\caption{\label{fig:DS_optimization}
		\textbf{Improving DS device geometry.} 
		\textbf{(A)} Device structure from Ref. [\textcolor{blue}{17}].
		\textbf{(B)} Improved device structure (DS$_{\textrm {imp}}$) 
		to reduce edge roughness for reflected electrons from second junction. 
		Ideally, the local gate should be restricted within the 
		triangle enclosed by green dash line to make the region free from edges and electrons can be redirected to the other source by second junction shown by white dash lines. 
		However, it is impossible to maintain smooth potential at the corners. Therefore, the left and bottom sides 
		are kept extended ($L_{ext}$ = 100 nm).
		\textbf{(C)} Comparison of on-off ratio of 
		DS vs. DS$_{\textrm{imp}}$. DS$_{\textrm{imp}}$ shows 
		less sensitivity to edge roughness.}
\end{figure*}

The main motivation behind DS device in Ref. [\textcolor{blue}{17}] 
was to use only junction in filtering procedure at the local gated region. 
In Fig. \ref{fig:DS_optimization}A, we can clearly see that this region 
is not free from edges and can be further optimized. 
Fig. \ref{fig:DS_optimization}B shows the structure 
(DS$_{\textrm{imp}}$) to improve the device performance 
in presence of edge roughness. $L_{ext}$ is kept to have feasible 
smooth electrostatics for both junction and reduce leakage path 
at the corners. Figure \ref{fig:DS_optimization}C shows 
comparison between DS and DS$_{\textrm{imp}}$ structure 
in terms of on-off ratio and we clearly see that 
DS$_{\textrm{imp}}$ is less sensitive to edge roughness.
\vspace{10mm}

\end{document}